\documentclass[pre,aps,pacs,twocolumn]{revtex4}
\usepackage{epsfig}
\usepackage{amssymb}
\usepackage{bm}
\usepackage{color}

\newcommand{\C}[1]{{\mathcal{#1}}}
\usepackage[latin1]{inputenc}
\newcommand{\beq}{\begin{equation}}
\newcommand{\eeq}{\end{equation}}
\newcommand{\bea}{\begin{eqnarray}}
\newcommand{\eea}{\end{eqnarray}}
\def\T{{\cal T}}
\def\TT{{\mathbb T}}
\def\GG{\cal G}
\def\OO{{\cal O}}

\begin{document}

\title{Ergodicity and Slowing Down in Glass-Forming Systems with Soft Potentials: No Finite-Temperature Singularities}

\author{Jean-Pierre Eckmann and Itamar Procaccia}
\affiliation{University of Geneva, 1211 Geneva 4, Switzerland\\The Weizmann Institute of Science, Rehovot 76100,
  Israel}
\begin{abstract}
The aim of this paper is to discuss some basic notions regarding generic glass
forming systems
composed of particles interacting via soft potentials. Excluding explicitly
hard-core interaction we discuss  the so called `glass transition' in which
super-cooled amorphous state is formed, accompanied with a spectacular slowing
down of relaxation to equilibrium, when the temperature is changed over
a relatively small interval. Using the classical example of a 50-50 binary
liquid of $N$ particles with different interaction length-scales we show that
(i) the system remains ergodic at all temperatures. (ii) 
the number of topologically distinct configurations can be computed, is temperature
independent, and is exponential in $N$. (iii)  Any two configurations in
phase space can be connected using elementary moves whose number is
polynomially bounded in $N$,
showing that the graph of configurations has the `small
world' property.
(iv) The entropy of the system can be estimated at any temperature (or energy),
and there is no Kauzmann crisis at any positive temperature. (v) The
mechanism for 
the super-Arrhenius temperature
dependence of the relaxation time is explained, connecting it to an entropic
squeeze at the
glass transition. (vi) There is no Vogel-Fulcher crisis at any finite
temperature $T>0$.
\end{abstract}

\maketitle
\section{Introduction}
It is not uncommon to read in papers devoted to the glass-transition a
statement of the type `it is well
known that glass forming systems lose ergodicity'. It is even more common \cite{96EAN} to
use fits of the
$\alpha$ relaxation-time $\tau_\alpha$ in such systems to the
Tamam-Vogel-Fulcher formula
\begin{equation}
\tau_\alpha \propto \exp\bigl(AT/(T-T_{\rm VF})\bigr) \ , \label{TVF}
\end{equation}
which indicates a belief that the relaxation time actually diverges at some
temperature $T_{\rm VF}>0$.
Related to these issues is the concept of the Kauzmann temperature \cite{48Kau} which is a
finite temperature $T_{\rm K}>0$ where the extrapolated entropy appears to vanish. In
this paper we argue that these
related questions should be addressed with care; we wish to clarify which of these
issues can actually appear in glass-forming systems, and which of them are only
a consequence of inadequate simulations or interpretations, or even of
a confusion
of questions.

In some parts this paper is a review, or an outgrowth of  our own efforts to
understand basic concepts
of glass-formation, which became more transparent in some of our earlier
work \cite{07ABHIMPS,07HIMPS,07Eck}, and whose findings we combine in
this paper. Glasses and their formation have occupied researchers for many decades, with ideas being
first developed on theoretical bases \cite{01Don}, and, in more recent years,
being studied extensively with the help of computer simulations \cite{89DAY,99PH,04DA,06ST}. Such
simulations vary from very realistic models to toy models with
simplified dynamics, but correspondingly faster simulations.
We can summarize our findings in the following logical structure:

1) One has to distinguish between large (but finite) systems as
compared to infinite systems. Infinite systems pose difficult
conceptual problems, because in this case, density and close packing are not
tightly related (any small lowering of density will allow for
arbitrarily large voids, and the jamming problems disappear).

We will therefore focus our discussion only on systems with a
finite, but arbitrary, number of particles. In this view an Avogadro
number of particles is finite, and it is important to state this fact.

2) There is an essential difference between systems or particles with a hard
core and systems of particles interacting with soft repulsive potentials. In the
first case, there is obviously a density where the particles
can not move any more (or at least not all of them can be moved). This
poses interesting problems of jamming, or contact geometry. These
issues have been studied in depth by Stillinger and his school, with a
careful analysis of different types of `movability' \cite{06Don}. We point out that
these problems, while very interesting from the geometrical point of
view (see also the book by Conway and Sloane \cite{Con}) do not really address the questions
of what a generic glass is. We will thus focus attention to systems with a finite number of
particles, and with soft potentials. 

3) There is another issue which deserves attention: time. Many concepts
of glasses are based on the notion that if something does not happen
before a given time, then it will never happen. Such reasoning is
unacceptable from a theoretical point of view, since a natural time
scale does not exist. We will argue that in systems with a finite number of
particles
interacting via soft potentials there is no singularity in the relaxation time
at any positive
temperature. In particular such systems undergo a glass
`transition' in the
sense that their relaxation time increases without limit as
temperatures are lowered, but they remain
ergodic. Observing
the consequences of ergodicity may necessitate waiting for an unbounded, but
still finite time. At any temperature larger than $T=0$ all of phase space is
available to the system, all the configurations are
dynamically connected; the configurational entropy can be computed, yielding a
finite result
for any given finite energy (or finite temperature). We stress and reiterate
that this paper deals with systems having a finite number of particles, which
are interacting via soft potentials and are being observed for an unbounded
time.

One can exemplify the discussion with the help of many simple models, and for 
concreteness we choose the
classical example of a glass forming binary mixture of particles interacting
via a soft $1/r^{12}$ repulsion with a `diameter' ratio of
1.4. More or less related models can be found in \cite{FA1984,davisonsherrington2000,sh2002,schliecker2002,Jstatmech2007,benarous2006}. This 2-dimensional model had been selected for simulation speed and, more importantly, for the ease of
interpretation. We refer the reader to the extensive work done on this system \cite{89DAY,99PH,07ABHIMPS,07HIMPS,07IMPS}. It shows that
it is a {\em bona fide} glass-forming liquid meeting all the criteria of a
glass transition. In short, the system consists of an
equimolar mixture of two types of particles with diameter $\sigma_2=1.4$ and
$\sigma_1=1$, respectively, but with the same mass $m$. The three pairwise
additive interactions are given by the purely repulsive soft-core potentials
\begin{equation}
u_{ab} =\epsilon \left(\frac{\sigma_{ab}}{r}\right)^{12} \ , \quad a,b=1,2 \ ,
\label{potential}
\end{equation}
where $\sigma_{aa}=\sigma_a$ and $\sigma_{ab}= (\sigma_a+\sigma_b)/2$. The
cutoff radii of
the interaction are set at $4.5\sigma_{ab}$. The units of mass, length, time
and temperature are $m$, $\sigma_1$, $\tau=\sigma_1\sqrt{m/\epsilon}$ and
$T=\epsilon/k_B$, respectively, with $k_B$ being
Boltzmann's constant.

\begin{figure}
\centering
\includegraphics[width=0.45\textwidth]{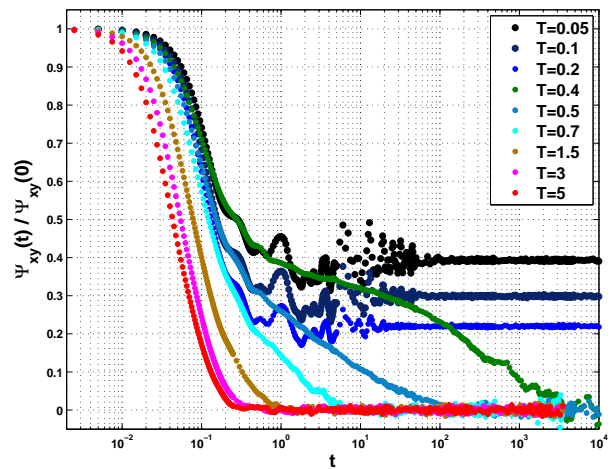}
\caption{(Color online:). Typical correlation functions at different temperatures for the model under discussion. Shown are the shear stress-shear stress correlation function which decays exponentially at high temperatures, and then, at about $T=0.5$, changes character to a stretched exponential form. At temperatures lower than 0.4 the correlations decay to a finite value, which is determined by the mean shear modulus, indicating that the material behaves like a solid for the given measurement time. For further details about these measurements see \cite{08IPS}.}
\label{stress}
\end{figure}

It has been shown that for  temperatures $T>0.5$ the system behaves like a liquid.
For temperatures lower than 0.5 the system begins to slow down, the correlation functions
stop decaying exponentially, and can be fitted within reason to a stretched exponential
form, cf.~Fig.~\ref{stress}. 
The time constant (or so-called $\alpha$ relaxation time) can be fitted to the Vogel-Fulcher
form (\ref{TVF}) for temperatures not too close to $T_{\rm VF}$. The model shows the expected behavior
of the specific heat at the temperature range that is considered the
`glass transition' ({\it i.e.}, where
the relaxation time increases rapidly), and the `super-Arrhenius' dependence of the viscosity
on the temperature. In short, the system appears as a good example
to consider to elucidate the fundamental issues that concern us here.

In Sect.~\ref{VOR} we explain how to discretize the configurational space using Voronoi tesselations
and their dual Delaunay triangulations. This offers the basis of the demonstration of ergodicity
at any finite temperature. We also review the mathematical results concerning the size of this
phase space (exponential in the number of particles $N$) and the fact that its graph has the small-world
property in the sense that any configuration can be reached from any other using a polynomial number
of steps. In Sect.~\ref{Conf} we discuss the configurational entropy, and explain that there is
never a Kauzmann crisis in this system. Finally in Sect.~\ref{slow} we present the fundamental reason
for the `super-Arrhenius' dependence of the relaxation time on the temperature. This is due
to the fast decrease in the concentration of some quasi-species, leading to an entropic squeeze.
In Sect.~\ref{disc} we summarize the paper and offer concluding remarks.

\section{Voronoi construction and ergodicity}
\label{VOR}

\subsection{Voronoi tessellation and Delaunay triangulation}

We begin by discussing the possible configurations of the system in a
systematic way, using
the time-honored Voronoi polygon construction
\cite{89DAY}. It associates with every configuration of the particles
a subdivision of position space into polygons, one per particle. These
polygons will also be called \emph{cages}.
The polygon associated with any particle contains all points
closer to that particle than to any other particle. The edges of such
a polygon are the perpendicular bisectors of the vectors joining the
center of the particle (actually the coordinates of the point
particles we consider). As had been noted in \cite{89DAY,99PH}, when
periodic boundary conditions are used, the average number of sides of the
polygons is exactly 6.  This follows because the Euler characteristic 
on the torus is 0: $\chi=V-E+F=0$, where $V$, $E$, and $F$ are
respectively the numbers of vertices (corners), edges and faces in the
given polyhedron. 

\begin{figure}
\centering
\includegraphics[width=0.30\textwidth]{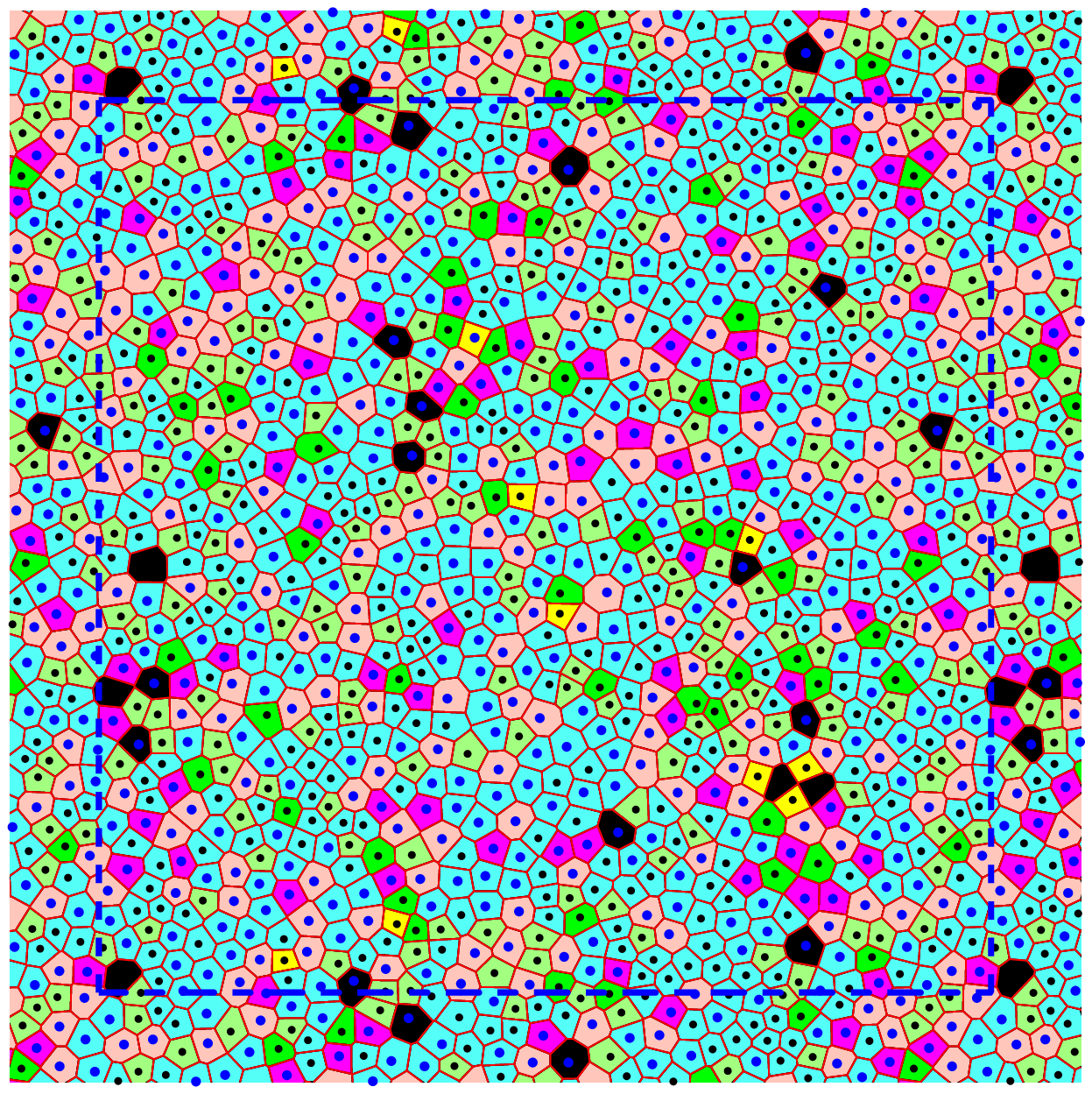}
\includegraphics[width=0.30\textwidth]{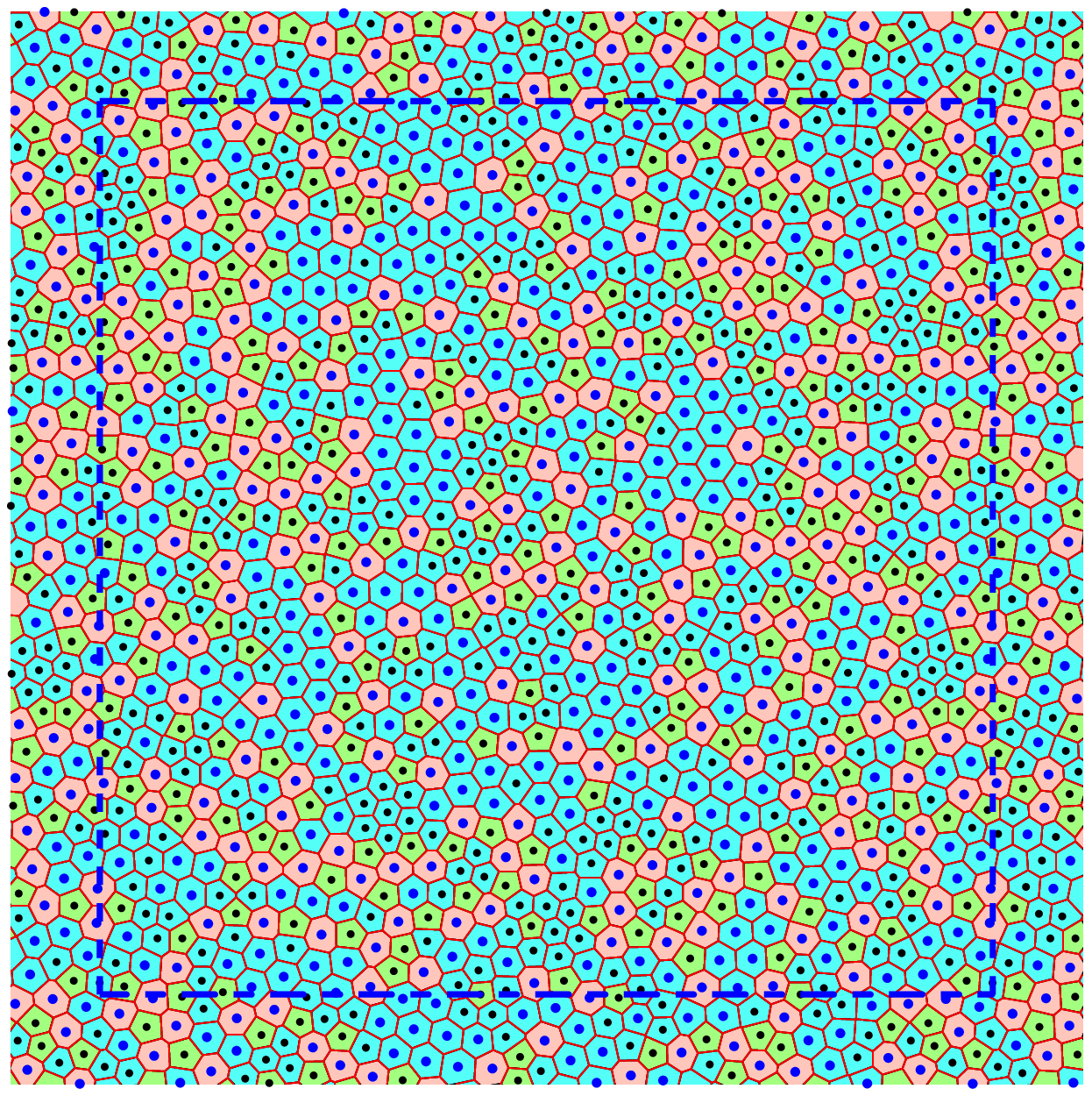}
\caption{Upper panel:  The Voronoi polygon construction in the liquid state at $T=3$  with the seven-color code used in this paper. Small particles in pentagons (heptagons) are light green (dark green) and large particles in pentagons (heptagons) are violet  (pink). Lower panel: a typical
Voronoi construction in the glass phase at $T=0.1$. Note the total disappearance of liquid-like
defects (large particles in pentagons and small particles in heptagons).}
\label{Vor}
\end{figure}

Typical such Voronoi tessellations for two different temperatures are shown in Fig.~\ref{Vor}.
Since there are two types of particles (small or large, or blue and red (respectively)), in order to have a mapping between the
particle positions and the number of sides in the polygons of the Voronoi tessellation, we
distinguish between polygons having a small and large particles in
their center. Thus, a coloring scheme of cells
will take into account not only the number of sides, but also the type
of the particle (big or small) in each cell. The tessellation obtained without distinguishing the type
of particles will be referred to as `colorless' and so, below, will be the resulting triangulation.
When distinguishing the two types of particles, the tessellation and its triangulation will be referred to as
colored.

Note that in a generic configuration the Voronoi decomposition has the
property that exactly three lines meet at a
point. Accordingly, the dual graph of such a decomposition is a
{\bf triangulation}. Such triangulations are called {\bf Delaunay triangulations}.
They are obtained by simply connecting the centers of
those particles which are separated by one edge of the polygon.
In other words, the vertices of
the triangles coincide with the particles at the centers of the
Voronoi cells. Let us remark that this construction can be also done
in 3-dimensions (or more), 
where now instead of edges of polygons we have faces of polyhedra and
four faces will meet  at a point. In this case, the dual graphs are now made of
tetrahedra. In 3-dimensions 
we will call the graphs obtained in this way Delaunay
tetrahedrization. It is noteworthy that there is  
an important difference between 2 and 3-dimension. In 2-dimensions,
the points of any triangulation can be arranged in such a way in
position space that the triangulation is indeed a Delaunay triangulation,
as described above, that is, the dual of a Voronoi decomposition.
However, in 3-d there exist
tetrahedrizations which are \emph{not} duals of Voronoi tessellations, or in
other words, cannot be obtained starting with particles spread in
3-dimensions, mapped onto a Voronoi tessellation and then tetrahedrized
(\cite{06santos}).

Having outlined the description of configurations in this model (and
in fact all models of this general type), we consider now the question
of dynamic accessibility, viz.~ergodicity.
The model under consideration is Hamiltonian, but in simulations it is coupled to a heat bath of
temperature $T$.
There are several methods for achieving such a coupling, leading to the study of this
model in either $N$, $V$, $T$ or $N$, $p$, $T$ ensembles.  In either ensemble the mean energy per particle is fixed, but there can be arbitrarily large fluctuations in the energy. It is precisely these fluctuations which are the source of ergodicity. The energy is partially kinetic and partially
potential. By ergodicity, one means that the natural motion of the
system can reach every point in phase space. For systems having one type of (indistinguishable) particles, it is not important to look at particles exchanging positions:  indeed  every point in phase space can be obtained by moving particles within their cages and translating cages as is necessary. In contrast,  when there is more than one type of particle (here we have two types),
ergodicity means that particles must also be able to change their relative
positions in all imaginable ways. It is therefore useful in both cases
to distinguish between thermal motion of individual
particles rattling in their own `cages', without really changing their
relative positions, and large scale movements and exchanges of particles that change
the macroscopic configuration. Only the second class of movements is
important in 
terms of the configurational entropy of the system, which arises from
counting different 
configurations, for which the thermal motion does not matter. For example, in systems with hard core
interactions, there are various levels at which motion can be
restricted, as we have mentioned before. But here, we consider
soft particles.  

Since we are starting from a problem with particles which live in a
concrete space,
we can focus on the restricted class of
Delaunay graphs (which actually coincides in 2-d with all
triangulations, as we have explained earlier).
In other words, in 2-dimensions (3-dimensions) the \emph{natural}
phase-space for our system is the set of all \emph{Delaunay} triangulations (tetrahedrization). 

\begin{figure}
\centering
\includegraphics[width=0.20\textwidth]{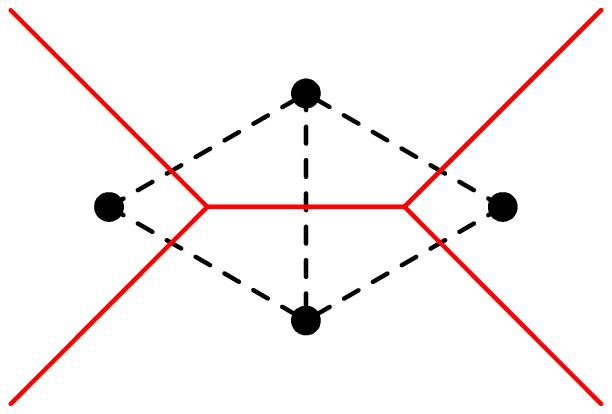}
\includegraphics[angle=90,width=0.20\textwidth ]{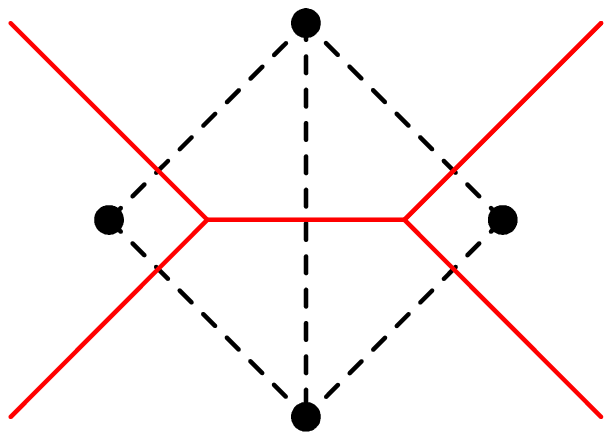}
\caption{ A T1 process in the Voronoi tessellation (in red lines) and the equivalent flip process
in the Delaunay triangulation (dashed curves)}
\label{T1}
\end{figure}

In 2-dimensions the elementary change in the Voronoi tessellation is obtained by a so called $T_1$
 process as seen in Fig.~\ref{T1}. (This operation has various names
 in the literature:
 Gross-Varstedt move, Pachner move, flip.)
 \cite{Malyshev1999,Negami1999,Mori2003,pach78,pach81}. Note that the operation of a
 $T_1$ process in the Voronoi  
 tessellation translates to a flip in the Delaunay triangulation, see Fig.~\ref{T1}.

To simplify things further, we consider instead of the triangulations of the
torus the triangulations $\T$
of the sphere. The simplification is that the genus is 0, and that
more is known about the combinatorics of triangulations of the sphere
than of the torus. Given $N$, corresponding to the number of particles
in the original model, we let $\T$
denote a triangulation of the sphere with $N$ nodes and we let $\TT_{N,0}$
denote the set of
all such triangulations. By this, one means the set of all combinatorially
distinct rooted simplicial 3-polytopes. In particular, a triangulation
should not have any `double edges'. We further refine the
definition, by distinguishing 2 types of nodes in the triangulation:
We first number the nodes from 1 to $N$ and then define
2 types of nodes.
Those with even index are the `small particles' and those with odd
index the `large' ones.  This means that the triangulation has the same
number of large and small nodes (up to a difference of one). We will
also call the two types of nodes two \emph{colors}.

\begin{figure}
\centering
\includegraphics[width=0.38\textwidth]{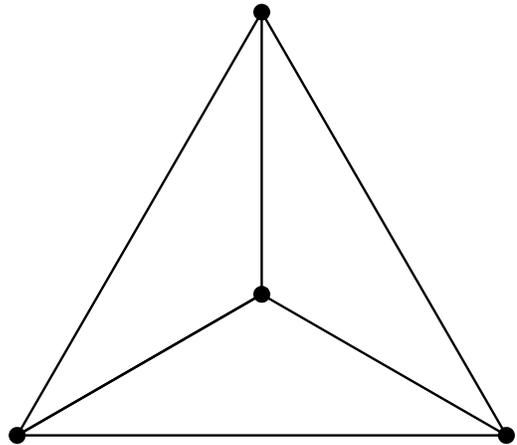}
\caption{A configuration with four nodes that cannot be flipped,
  because that would lead to double links. In 3-dimensions there is
  more than one corresponding graph, but none of them is the dual of a
Voronoi \cite{06santos},}
\label{stuck}
\end{figure}

 We shall call the odd nodes \emph{blue} and the even ones \emph{red}
  and will refer to the triangulations as \emph{colored} triangulations.
Once the colors are assigned, the numbers are again forgotten.
The set
of all colored triangulations with $N$ nodes will be called $\TT_{N}$.
This is our \emph{phase space} and the dynamics is mapping points in this
phase space to other points.  Having understood that, we can now ask the major question of this section:

{\bf The question of ergodicity}: is every point in phase-space
accessible through a sequence of motions that are flips. The answer is
not obvious from the outset. For example in 2-dimensions one can
consider a triangulation that includes locally a form as in
Fig.~\ref{stuck}. 
Obviously, a form like this is stuck, since one
cannot do any flip 
without doubling one of the links. (One can see this easily by noticing
that the configuration in Fig.~\ref{stuck} is a valid triangulation of the
sphere, and it is the unique triangulation with four points.) The
question of ergodicity is even less trivial in 3-dimension, where
there is a whole class of unflippable tetrahedrizations.
However, they are not duals of
Voronois. But, the relevant question (in any dimension) is whether the
triangulations (tetrahedrizations,\dots) which {\em are duals of
  Voronoi tessellations}  are indeed all connected in one large component.
The answer is nevertheless `yes' in both 2 and 3-dimensions as we explain next.

\subsection{Demonstration of ergodicity}

One can attack this problem in two ways, the first being more
physical, and the second more mathematical. The physical argument is
trivial. Since the particles are soft, and the energy is not bounded,
they can be moved around each other in any way one can imagine. 
We stress again that while this needs perhaps a lot of energy, the
large fluctuations of the energy will guarantee that this will
eventually happen, rarely, but surely.
And hence ergodicity is obvious. The only care one must observe is that
moves through degenerate situations (4 lines meeting at a Voronoi
vertex) must be avoided. 

\begin{figure}
\centering
\includegraphics[width=0.30\textwidth]{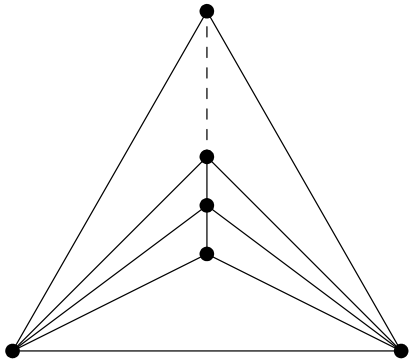}
\caption{The Christmas tree. Two nodes are at the bottom, the others
  (only 4 shown) are in the stem of the tree.}
  \label{Xms}
  \end{figure}

The question is more intriguing when formulated in terms of
triangulations alone, since we have seen in Fig. \ref{stuck} that a local configuration in which  3
links emanate from a node cannot be flipped. Could it be that the triangulation
is so imbricated that in fact none of the edges can be flipped?
Indeed, for 4 particles this is exactly what happens, but then, the
phase space consists of exactly one triangulation and thus there is no
need to move any link. In 1936, Wagner \cite{Wagner1936} showed that a
finite number of flips will transform any triangulation of the sphere into a
`Christmas tree,'
which is the configuration shown in Fig.~\ref{Xms}. See also
\cite{CE2005} for a discussion of several related issues.
Since one can
undo the flips, this implies immediately that any two triangulations
can be connected by a sequence of flips going through the christmas tree. There is abundant literature on this question \cite{06santos, Negami1999} which also plays a certain role in the
classification of 3-manifolds. 

\subsection{The size of phase space}

 The possible states of our system of triangulations with $N$ nodes is the set $\TT_{N}$
of all possible
colored triangulations. The set $\TT_N$ has, as we will see, a number of
elements
which grows like $C^N$ for some constant $C>1$. It is thus a discrete
space with a finite number of states. To describe the dynamics of
flipping in a geometric way, one should view this set as the
{\em dynamical graph}  $\GG$, whose nodes are now the elements of the set $\TT_{N}$ (not to
be confused with the nodes (particles) of any triangulation \/$\T$) and two
of its nodes are linked if one can be reached from the other by a
flip. (This makes an undirected graph, since one can flip back and
forth.) The reader should note that there are two graphs in this
discussion: Each triangulation is a graph with $N$ nodes, and $3N-6$
links (by Euler's theorem), while the graph $\GG$ has about $C^N$
nodes, and about $3N-6$ links \emph{per node}. This last statement
follows because in every state of $\TT_{N}$, one can choose which of the
$3N-6$ links of the triangulation $\T$ one wants to flip. However,
there will, in general, be somewhat fewer links which are candidates
for flipping, because whenever there is a node of degree $3$ in the
triangulation $\T$ its links can not be flipped (a tetrahedron is
unflippable).

Finally, given any two elements in $\TT_{N}$, that is, any two
triangulations with $N$ nodes, we will show below that $\OO(N^2)$ flips
are sufficient to walk on the graph $\GG$ from one to the other. Thus,
 the diameter of the graph $\GG$ is at most $\OO(N^2)$ while it has
$\OO(1)^N$ vertices. This means that $\GG$ has the `small-world' property
 \cite{Watts1998}. It has also small clustering coefficient, since
 there are very few triangles in the graph $\GG$ (it is difficult to
 get from a triangulation back to the same triangulation with 3 flips).

In the remainder of this section, we spell out these statements. They are
well-known for uncolored graphs, so the only task is to prove them for
the colored graphs, see \cite{07Eck}.

We first state two known results for the set $\TT_{N,0}$ of uncolored
triangulations: \\
{\em Lemma 1} \cite{Tutte1962,Negami1999,Mori2003}
  The number of elements in $\TT_{N,0}$ is asymptotically
  \begin{equation}\label{tutte}
\left(\frac{256}{27}\right)^{N-3} \frac{3}{16 \sqrt{6\pi N^7}}~.    
  \end{equation}
The distance between any two uncolored triangulations is 
at most $6N-30$ flips.

For the case of the colored graphs, with $N_{\rm red} = N_{\rm blue}
+c$
and $c\in\{0,1\}$, that is, about equal number of red and blue nodes,
one has\\
{\em Lemma 2}\
  The number of elements in $\TT_N$ is asymptotically bounded above by
  \begin{equation}\label{tuttecolor}
2^N \left(\frac{256}{27}\right)^{N-3} \frac{3}{16 \sqrt{6\pi N^5}}~,
  \end{equation}
and below by the expression (\ref{tutte}).
The distance between any two colored triangulations in $\TT_N$ is bounded by
\begin{equation}\label{diam}
  C_1 N^2 + C_2
\end{equation}
flips with some universal constants $C_1$, $C_2$.

We note that the phase space as defined here is (obviously) independent of the temperature. We can thus conclude that the present classical model of glass-formation does not suffer
from any issue of loss of ergodicity. Accordingly, it should have a valid statistical mechanics
at any temperature $T>0$. Next we show that indeed its configurational entropy $S$ never
suffers any finite temperature crisis, and there is no Kauzmann temperature where $S\to 0$

\section{Configurational Entropy}
\label{Conf}

\subsection{Statistical Mechanics}

Needless to say,  the configurations discussed in the previous section have different energies
and therefore the probability to see any particular one can be strongly dependent on the temperature. 
To discuss the temperature dependence of the configurational entropy of this system we need to review the  statistical mechanics that was introduced for this system in \cite{07HIMPS}. The basis of the analysis
is again the Voronoi tessellation.

\begin{figure}
\centering
\includegraphics[width=0.38\textwidth]{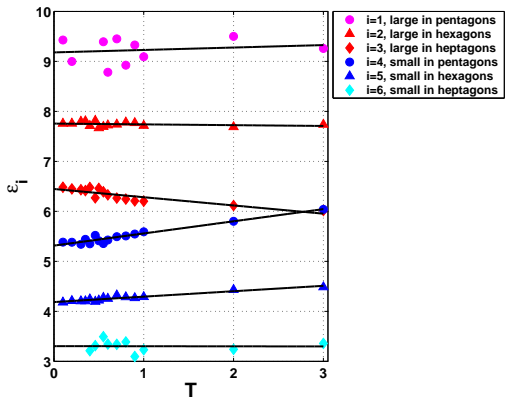}
\caption{(Color online). The average energies of the Voronoi cells as a function of the temperature as measured in the simulations.}
\label{energies}
\end{figure}

To get some systematics, we first define a `typical' energy for each
type of topological cell, with an average taken over all cells with a
given number of sides and a given particle type (big or small) in its
center. The average is then
 $ \epsilon_i=\langle\, \sum_{k=1}^{E_i}\epsilon \left(\frac{\sigma_{ik}}{r_{ik}}\right)^{12}\rangle $ where $E_i$ is the
 number of edges associated with that cell type ($i$), and
$r_{ik}$ being the distance to the particle in the adjacent Voronoi
 cell and the average is over all particles of the same type $i$. In
Fig.~\ref{energies} we present the values of these energies measured
numerically
as a function of the temperature, following a protocol of slow cooling. In the
range of temperatures
explored by simulations there are 10 different cell types, (large
particle or small particle in squares, pentagons, hexagons, heptagons
and octagons), but octagons and squares are not shown since they already
disappear at relatively high temperature,  much above the glass
transition.

We learn from these data graphs that the different cell
types have clearly 
split energies throughout the interesting temperature range, and that
these energies are only 
weakly dependent on the temperature. Within the temperature range of
interest we can focus on the six types of cells; denote by
$\{N_i\}_{i=1}^6$ the number of cells of each type, with number of
edges $E_i$, ordering them by the mean energy $\epsilon_i$ with $i=1$ being
the highest  (large particle in a pentagon) and $i=6$ being the lowest 
(small particle in an heptagon). Additional important properties of the
cell types are their areas $\Omega_i$ and their shapes; the first
affects the enthalpy term  and both affect the configurational entropy when we count
the number of possible tilings of the plane. 
\begin{figure}
\centering
\includegraphics[width=0.38\textwidth]{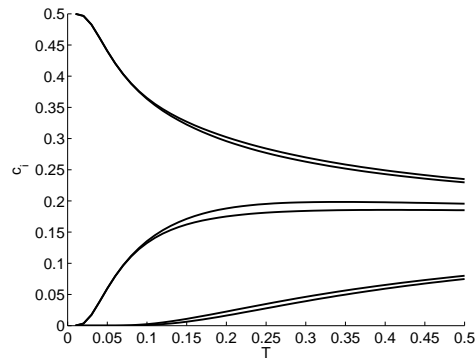}
\caption{The concentrations of all the Voronoi cells as a function of $T$. Note the close-to-degeneracy of pairs of cells: the upper pair are the glass-like quasi-species, the middle pair are the two hexagons and the lower pair the liquid-like quasi-species. Observe the existence of two typical temperatures ranges, one around $T_2$ where the liquid-like quasi-species deplete severely, and the
other around $T_1$ where the hexagons deplete rapidly.}
\label{statmech}
\end{figure}

With this in mind we can construct the statistical mechanics of this system by considering
the free energy $G=U+pV-TS$.  We should stress at this point that one
could aspire for more accurate statistical mechanics, considering for
example not only the type of particle inside the Voronoi n-gon, but also who
are the neighbors  (small or large particles)
(which resembles somehow the plaquette expansions in statistical mechanics).
Such a choice would have allowed a better treatment of the tendency of
hexagons with large (or small) particles for example to crowd together to minimize the $pV$ term in the
free energy. The price is that the number of quasi-species 
increases to 42 (6, 7 and 8 for each pentagon, hexagon and heptagon respectively). While doable and a bit more precise, this more involved statistical mechanics does
not shed more light on the issues of principle that interest us in the present paper, and therefore we
do not discuss such improvements any further. 

Coming back to the simpler variant, we note that in the free energy
the value of $U$ is
$\sum_{i=1}^6 N_i \epsilon_i$. The $pV$ term is simply $p\sum_{i=1}^6 N_i
\Omega_i$.
 Lastly, we need to estimate the entropy term. In principle this should be
computed from the number
 of possible complete tilings of the area by $N_i$ cells of each type with its
given area
 and shape, subject to the Euler constraint $\sum_{i=1}^6 N_i E_i =6N$, where $E_i$
 is the number of edges of the $i$'th polygon. This is
a formidable problem. A useful estimate can be obtained by considering the area
 only, and filling space starting with the largest objects, then the
 next largest, and so on, until the smallest 
 are fit in. We do this by dividing the (remaining) volumes into
 boxes, and studying the combinatorial filling of these boxes.
 Denoting the possible number of boxes to fit the largest cells by ${\cal
N}_1\equiv V/\Omega_1$, then the number of boxes available for the second
largest cell by $\C N_2\equiv
 (V-N_1\Omega_1)/\Omega_2$ etc., the number of possible configurations $W$ is
 \begin{equation}
 W = \prod_{i=1}^6 \frac{\C N_i!}{N_i!(\C N_i-N_i)! }\ . \label{W}
 \end{equation}
Using the abbreviation $x_i\equiv N_i/\C N_i$ we compute directly $x_i=c_i\Omega_i/\sum_{j=i}^6 c_j\Omega_j$
where $c_i$ is the number concentration of each defect. We can now compute $S=k_{\rm B}\ln W$ and write $G$ together with a Lagrange multiplier for the Euler constraint,
\begin{eqnarray}
&&G = \sum_{i=1}^6 N_i \epsilon_i + p\sum_{i=1}^6 N_i \Omega_i +\lambda \sum_{i=1}^6 N_i E_i  \nonumber\\
&&+T\sum_{k=1}^6 \C N_k[x_k \ln x_k+(1-x_k)\ln (1-x_k)]  \ . \label{G}
\end{eqnarray}
The chemical potential $\mu_i \equiv \partial G/\partial N_i$ is then,
for $i=1,\dots,6$,
\begin{equation}
\mu_i =\epsilon_i +p\Omega_i +T[\ln x_i +\sum _{k=1}^{i-1} \frac{\Omega_i}{\Omega_k}\ln(1-x_k)]+\lambda E_i \ . \label{general}
\end{equation}
We now recognize that Voronoi cells of different values $E_i$ but with the same size particle
(small or large) are in equilibrium, each one being able to change to another, but small particles
cannot change to large particles, and therefore in equilibrium there exist only two independent values of
$\mu_i$, one
for the small particles $\mu_S$  and one for the large particles $\mu_L$, and
we have 9 unknowns --
six values of $c_i$, 2 values of $\mu$ and one Lagrange multiplier $\lambda$.
This is precisely
balanced by the 6 equations (\ref{general}), the Euler constraint, and the two
constraints $\sum_{i=1}^3 c_i=\sum_{k=4}^6 c_k=1/2$. These equations could be
solved numerically
using the precise values of $\Omega_i(T)$ and $\epsilon_i(T)$ as measured in
the simulation.
The approximate calculation of the entropy however does not warrant such a
detailed calculation.
In reality, calculating the average areas of the cell types in the numerical
simulations, we discover
that to an excellent approximation these fall in two classes, smaller cells of
area $\Omega_S$ when
small particles are in them, and larger cells of area $\Omega_L$ where large
particles are enclosed.
These areas again are only weakly dependent on the temperature. Then the whole
system of
equations simplifies to two analytically tractable sets of equations
\begin{eqnarray}
\tilde \mu_L &=& \epsilon_i + T\ln c_i +\lambda E_i \ , \quad \{i=1,2,3\} \ , \nonumber\\
\tilde \mu_S &=& \epsilon_i + T\ln c_i +\lambda E_i \ , \quad \{i=4,5,6\} \ , \label{simple}
\end{eqnarray} 
together with the above mentioned three constraints. In $\tilde \mu$ we have absorbed terms that
added to $\mu$ in this special case. 

In Fig.~\ref{statmech} we show the solutions of these equations
when we use values of $\epsilon_i$ taken from the Fig.~\ref{energies} at $T=0$.  We learn from these results that the statistical
mechanics predicts that the first quasi-species to disappear are the
large particles in pentagons 
and the small particles in heptagons. While the first is the highest
in energy, the disappearance in tandem of the second is a result of
the Euler constraint, and could not be guessed {\it a priori}. In previous 
work the first disappearing quasi-species were called `liquid-like', since they are common in the
liquid state and their concentration is exponentially small in the glass state. The region of temperature
where their concentration falls off rapidly was identified with the region of slowing down. In fact, 
in \cite{07HIMPS} a quantitative relation between the concentration of these liquid-like quasi-species
and the relaxation time was derived, explaining the slowing down as a result of an entropic squeeze. We will return to this issue in Sect.~\ref{slow}.

\begin{figure}
\centering
\includegraphics[width=0.300\textwidth]{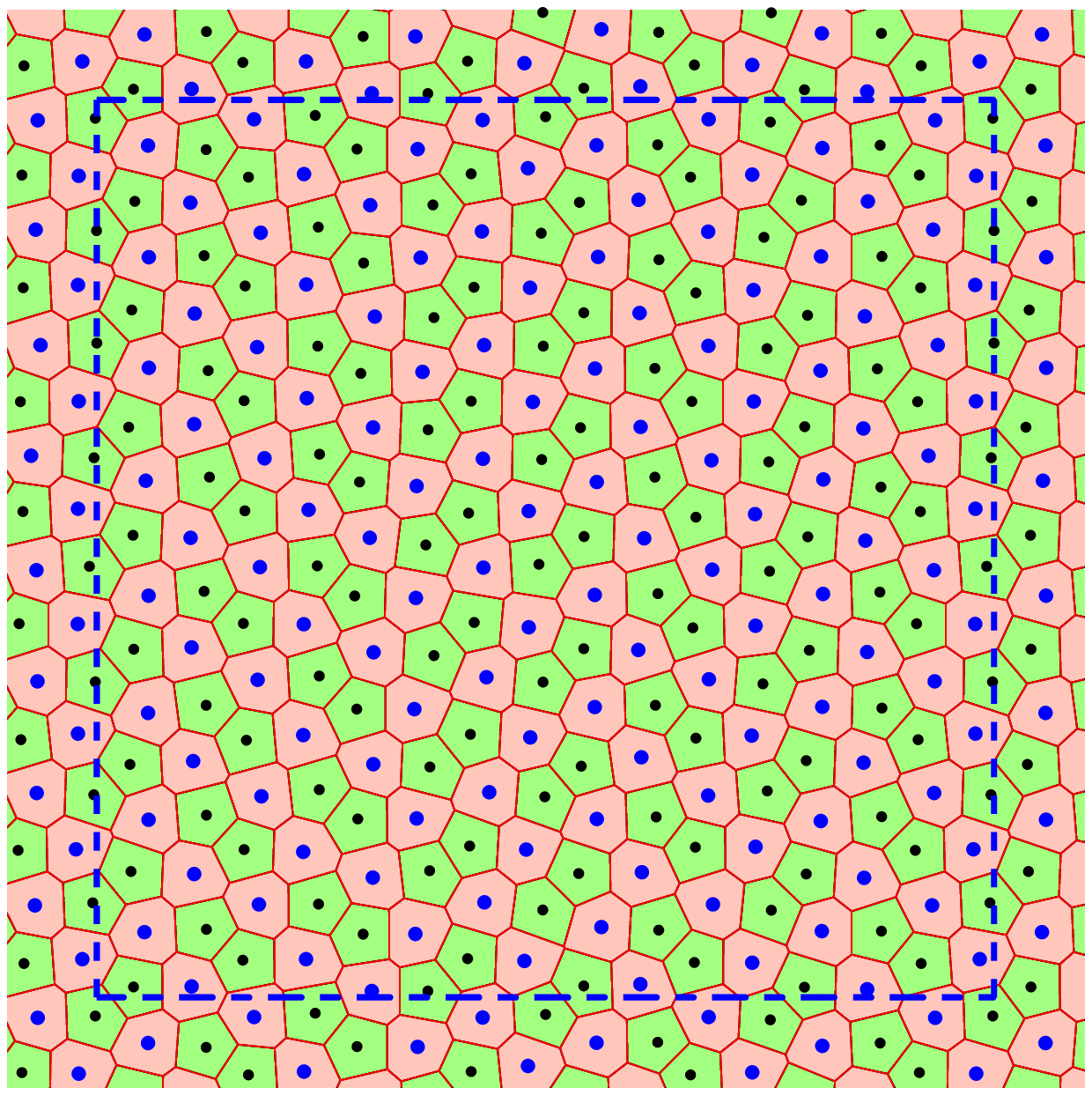}
\caption{(Color online). An example of a stable phase formed by glass-like quasi-species at very low temperatures.}
\label{crystal}
\end{figure}

The statistical mechanics predicts a second transition
(cf.~Fig.~\ref{statmech}). Below some temperature the concentration of
hexagonal cells is predicted to be exponentially small, and the system
retains only pentagons with small particles and heptagons with large
particles. These are referred to as the `glass-like'
quasi-species. Indeed, in \cite{07HIMPS} it was found that a phase made of
only pentagons with small particles and heptagons with large particles
exists and is stable at low temperatures, see Fig.~\ref{crystal}. 
Upon warming up such a phase, at a temperature roughly around $T_1$, a sizable
number of hexagons appears to form the generic
glassy state. Upon further warming, crossing a temperature roughly around $T_2$, a sizable number 
of liquid-like defects brings the system to a liquid state. The actual values
of $T_1$ and $T_2$ can be understood from this model. Denote by $c_\ell$, $c_H$ and $c_G$ the concentrations of the liquid-like, hexagons and glass-like quasi-species, and by $\epsilon_\ell=\epsilon_1+\epsilon_6\approx 12.48$ as the energy associated with the liquid-like quasi-species, by $\epsilon_H=\epsilon_2+\epsilon_5\approx 11.94$ the energy of the
hexagons, and $\epsilon_G=\epsilon_3+\epsilon_4\approx 11.76$ the energy of the glass-like defects. The theory predicts that ratios $c_\ell/c_H$ and $c_H/c_G$ are of the order of $\exp[-(\epsilon_\ell -\epsilon_H)/T ]$ and 
$\exp[-(\epsilon_H -\epsilon_G)/T ]$ respectively. As an estimate of $T_2$ and $T_1$ take these ratios to be, say,
of the order of $ 1\%\sim \exp(-5)$ and observe that such ratios are obtained for $T=T_2\approx 0.11$ and $T=T_1\approx 0.04$. It is important to notice that $\epsilon_H -\epsilon_G$ could be positive rather than negative, and then the system would crystallize on a hexagonal lattice. Such a lattice can exist in this system only when the particles phase separate into two pure hexagonal lattices of small and large particles respectively, with an interface in between. Such a phase may even be the ground state, but seems to be inaccessible in dynamical experiments starting from random organizations of small and large particles. 

\subsection{Nonexistence of Kauzmann temperature}

Assuming that indeed the realized state at $T=0$ is the state shown in Fig.~\ref{crystal},
but equivalently if the ground state were made of hexagons, it is obvious now that any finite temperature will allow the appearance of the other quasi-species (hexagons inside the phase in Fig.~\ref{crystal} or
pairs of pentagons and heptagons in the hexagonal phase if the latter were the ground state).
Focusing on the first situation, we understand that any pair of hexagons costs a given amount of
energy $\epsilon_H$ which is a given $\Delta$ above the ground-state $\epsilon_G$. 
The calculation given by Eq.~(\ref{G}) is legitimate at any temperature $T$, and the configuration
entropy is approximately correct as stated there. One can improve the calculation of the entropy
compared to the approximation employed above, but there is nothing extraordinary that is expected
at any temperature. We thus state that the configurational entropy is expected to be an analytic
function of $T$ at any value of the temperature, and it can vanish
only at $T=0$. There is no Kauzmann temperature here or in any similar generic model.

\subsection{The notion of fictive temperature}

Notwithstanding all of the above, the system under study can slow down so
much that upon reducing 
the temperature one has to wait for a very long time before
equilibrium is reached. When the relaxation times are already very
large, say at an initial temperature $T_{\rm i}$, any rapid decrease in the
temperature of the heat bath to a final temperature $T_{\rm f}$ may result in a very
lethargic response of the system, which may keep the  
concentrations of various quasi-species at values which are consistent
with $T_{\rm i}$ rather than 
$T_{\rm f}$. It is then perfectly legitimate to introduce the notion of a fictive temperature $T_{\rm fic}\approx T_{\rm i}$, as long as one is satisfied with short observation times. For longer and longer times the system
will exhibit the process of aging, and in particular $T_{\rm fic}$ will converge to $T_{\rm f}$, reaching there with certainty if given enough time. 

\section{Slowing down and entropic squeeze}
\label{slow}

The aim of this section is to explain the most important aspect of
glass formation, {\it i.e.},
the extreme slowing down in relaxation to equilibrium when the temperature is lowered.
The riddle is as follows: the natural time scale is determined by the molecular jitter due
to thermal motion. This time is typically of the order or $10^{-12}$s at room temperature.
Glassy dynamics exhibits relaxation time $\tau_e$ of the order of seconds, or hours, sometimes years.
How is it that such a huge gap in time scale is obtained {\em without} geometrical obstruction
(as is the main theme of this paper)? We will explain that the issue is entropic squeeze,
or the failure of entropy or `the number of available paths'  to overcome the necessary
energy climb required for relaxation.

In \cite{07ABHIMPS,07ILLP} it was argued that the relaxation time can be predicted if one knew
the typical scale $\xi$ that separates `liquid-like' quasi-species. In
other words, having the concentration  $C_\ell$ of large particles in
pentagons and small ones in heptagons one introduces a typical scale
$\xi$ by
\begin{equation}
\xi \sim \frac{1}{\sqrt{C_\ell}} \ , \label{xi}
\end{equation}
since the system has two dimensions.  To connect between the relaxation time $\tau_e$ and the length scale $\xi$ it was asserted  that for the viscous fluid there exists a free energy of activation $\Delta G^*(T)$ associated with the relaxation event,
\begin{equation}
\tau_e = \tau_0 \exp (\Delta G^*(T)/T) \ ,
\end{equation}
where $\tau_0$ is a microscopic time scale of the order of a single
particle vibration time. The free energy of activation is estimated as
the number of Voronoi cells $N^*(T)$ involved in the relaxation event,
times the (temperature independent) chemical potential per cell
$\Delta \mu$, $\Delta G^*(T)\approx N^*(T) \Delta \mu$. The number
$N^*$ depends on whether the relaxation event
is a 1-dimensional \cite{07ILLP} or 2-dimensional event \cite{07ABHIMPS}. In the first case $N^*(T)\approx \xi(T)/ \sqrt{\bar {\Omega}}$
while in the second $N^*(T) \approx \pi \xi^2(T) /4 \bar {\Omega}$, where $\bar {\Omega}$ is the mean area of a Voronoi cell. We end up with the predictions
\begin{eqnarray}
\tau_e&=& \tau_0 \exp ( \xi(T)\Delta \mu /\sqrt{ \bar {\Omega}} T) \ , \text{1-d event} \ , \label{tau1}\\
\tau_e &=& \tau_0 \exp (\pi \xi^2(T)\Delta \mu / 4\bar {\Omega} T)\ ,  \text{2-d event} \ . \label{tau2}
\end{eqnarray}
These predictions were shown to fit the simulation data very well \cite{07ILLP,07ABHIMPS}. Here we want to explain the fundamental origin of these formulae.

\begin{figure}
\centering
\includegraphics[width=0.300\textwidth]{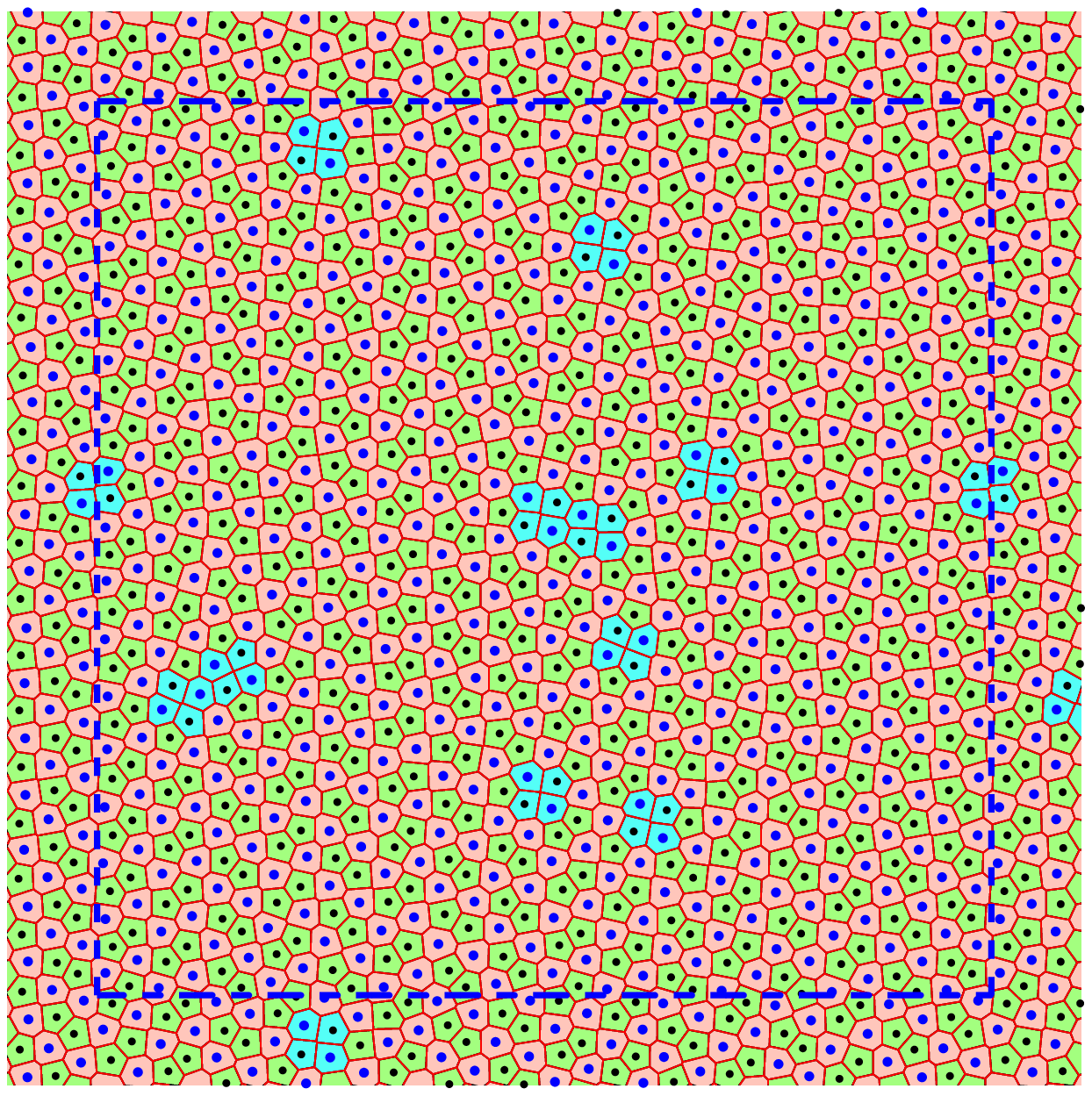}
\caption{(Color online). An example of a phase obtained by simulation at low temperatures when
hexagons appear in the phase of Fig. \ref{crystal}. Note the tendency of hexagons to clump together,
a tendency that is ignored in our approximate estimate of the configurational entropy.}
\label{hexa}
\end{figure}

The glass transition and the associated slowing down take place in the range of temperatures around
$T_1$  where the liquid-like quasi-species deplete rapidly. It is advantageous theoretically
to focus on the range of temperatures around $T_2$ where the hexagons deplete quickly, cf.~Fig.~\ref{hexa},
since then we have a smaller number of quasi-species to take into
account, while the fundamental phenomenon of entropic squeeze is not
very different. So think about a situation when the majority of
quasi-species are pairs of glass-like pentagons and heptagons, and set
up the energy scale such that these pairs have energy zero. Next consider a temperature $T$ where the equilibrium concentration $c$ of hexagons is small, $c\ll 1$. Set up the energy units such that each pair of hexagons (one with small particle and one with large particle) has an energy $\Delta=1$. Accordingly the energy of
this configuration is
\begin{equation}
E = cN \ . \label{ene}
\end{equation}
One the other hand, the entropy of this configuration can be estimated as the logarithm of
the number of ways the hexagons can be distributed, which is 
\begin{equation}
S\approx -Nc \ln c =Nc\ln(1/c) \ .
\end{equation}
As a result we can compute 
\begin{equation}
\frac{\partial S}{\partial E} =\frac{1}{N} \frac{\partial S}{\partial c} =-\ln c-1\approx -\ln c~.
\end{equation}
Remembering the thermodynamic identity $\partial E/\partial S=T$ we then conclude 
that the concentration of hexagons satisfies the relation
\begin{equation}
c = \exp(-1/T) \ . \label{c}
\end{equation}
Accordingly we conclude that in two-dimensions the average distance $\xi$ between hexagons, which  is $\xi\approx 1/\sqrt{c}$, satisfies
\begin{equation}
\xi \approx  \exp(1/2T)  \ . \label{xi2}
\end{equation}

Whenever a pair of hexagons
is created, with high probability, this will be undone sometime in
the future (by the inverse operation). Once
they are separated in space by the distance $\xi$, we need a number of flips which is
at least of the order of $\xi$, but maybe many more,  in order to annihilate a pair. This is the fundamental process of relaxation at temperature $T$ which we now proceed to estimate. In other words, we estimate how many flips are typically needed in order to get rid of one pair of hexagons, and we will show that
the answer is super-Arrhenius. 

We first note that because the concentration $c$ is small, any random flip will create a pair
of hexagons and increase the energy by unity. Only flips that eliminate a pair of hexagons (to create a heptagon and pentagon) reduce the energy by unity. We denote the probability of such a rare
event by $\gamma$. {\emph {It will be crucial that $\gamma$ depends on
  the temperature as we will explain below}.} Out of all the other flips, assume that a fraction $\alpha$ of flips does not change the energy (it has to involve a hexagon and one of a pentagon-heptagon pair). What is then the best
way to move one hexagon a distance $\xi$ until it can annihilate its counterpart? Assume that the
path takes $m$ flips where the energy increases by one unit (a typical flip),  is energy neutral
over $k$ flips, and  then goes down in energy in $m$ steps, with the constraint
\begin{equation}
2m+k\ge \xi \ . \label{cons}
\end{equation}
Of course, these events can take place in any order.
Clearly, we have a competition between the number of ways to arrange such a path and the 
energy barrier that needs to be surmounted, {\it i.e.},  we need to
sum up over all $m$ and $k$ the expression
\begin{equation}
P(m,k) = \frac{(2m)!} {m!\, m!\, k!} (1-\alpha-\gamma)^m\alpha^k \gamma^m e^{-m/T} \ .
\end{equation}
Note that all the factors are smaller than 1 and therefore the sum is
well approximated by the largest term, which occurs for $m=\xi/2$ and
$k=0$.
We thus get
\begin{equation}
P =\sum_{2m+k\ge \xi}P(m,k)\sim [(1-\alpha-\gamma)\gamma e^{-1/T}]^{\xi/2} \ .
\end{equation}
Note now that $\alpha$ is just some number smaller than 1, but more
importantly, $\gamma$, which is the probability to find an
energy-lowering move is proportional to the density of
defects. Indeed, if any flip is done at a link in the triangulation in
which neither of its ends is a defect
 the energy will go \emph{up}. (In fact, this reasoning also shows
 that $\alpha $ goes to zero when the temperature goes to 0.)

Now, the density of defects is $c=\exp(-1/T)=\gamma$ so that we get
finally,
$$
P\sim  (e^{-1/T} e^{-1/T})^{\exp(1/2T)/2} \ .
$$
The time we need to wait to see the event is proportional to the inverse of this probability, or
\begin{equation}
\tau_e \propto \exp{(\xi/T)}
\end{equation}
which is precisely Eq.~(\ref{tau1}). This is the one appropriate for 1-dimensional relaxation events
as assumed here and the generalization to other dimensions is obtained
by modifying the relation between the density and $\xi$. The
non-Arrhenius nature of the relaxation time is due to the strong
temperature 
dependence of $\xi(T)$ (cf.~\cite{07ILLP}), which in turn is due to the fast disappearance of 
a class of quasi-species. This reduction of the number of quasi-species is responsible for the entropy squeeze.

\section{Summary and Conclusions}
\label{disc}

Glass forming systems with hard cores can get jammed because of geometric constraints.
In this paper we argued, on the basis of a generic example,  that when
the potentials are soft, the spectacular slowing down associated 
with the glass transition is in a sense more interesting, since it does not occur due to geometric jamming.
Such system never lose their ergodicity, and any configuration can be
reached from any other in a polynomial (in $N$) number of steps, even
though the number of configurations is exponential in the number of
particles. We demonstrated explicitly that the configurational entropy
in such systems 
if finite at any temperature, and thus neither the Kauzmann temperature nor the Vogel-Fulcher formula
can be taken seriously. Both are the result of an extrapolation which is not fundamental. Finally we addressed the question of what is the reason for slowing down, and what is the mechanism
of its super-Arrhenius temperature dependence. We showed by an explicit calculation for a generic
relaxation step that near the glass transition when the concentration of some quasi-species becomes very small, the entropic squeeze results in the inability of the entropic counting of paths to balance the
energetic barriers, leading therefore to relaxation times that depend on the temperature
much faster than expected from the Arrhenius form.

In summary, we focused on the topological properties of generic glass forming systems, to clarify
some fundamental issues which are not always clear in the literature. Needless to say, much of the interest in
glass forming systems, including their mechanical properties, calls for understanding further issues,
including metric issues that are outside the scope of this paper. For some recent thoughts on these
subjects we refer the reader to \cite{07IMPS,08IPRS}.

{\bf Acknowlegdments}. We thank F.~Santos for many helpful discussions
on the issues of flips. We also thank D. Mukamel for drawing our
attention to related work. The work of JPE was partially supported by the
Fonds National Suisse and the Joseph Meyerhoff visiting professorship. IP is supported in part by the German-Israeli Foundation, the Israel Science Foundation and the Minerva Foundation, Munich, Germany.

\end{document}